\documentstyle[pre,aps]{revtex} 
\input{epsf}
 
\begin{document} 
\title{Asymmetries in Structure Factor Histograms} 
\author{G. Korniss, B. Schmittmann and R. K. P. Zia} 
\address{Center for Stochastic Processes in Science and Engineering \\ 
and \\ 
Department of Physics, Virginia Polytechnic Institute and State \\ 
University \\ 
Blacksburg, Virginia 24061-0435, USA} 
\date{February 26, 1997} 
\maketitle 
 
\begin{abstract} 
We investigate the dynamics of a three-state stochastic lattice gas, 
consisting of holes and two oppositely ``charged'' species of particles, 
under the influence of an ``electric'' field, at zero total charge. 
Interacting only through an excluded volume constraint, particles can hop to 
nearest neighbour empty sites. Using a combination of Langevin equations and 
Monte Carlo simulations, we study the {\em probability distributions} of the 
steady-state structure factors in the {\em disordered} phase where 
homogeneous configurations are stable against small harmonic perturbations. 
Simulation and theoretical results are in excellent agreement, both showing 
{\em universal} asymmetric exponential distributions.
 \\
 
\end{abstract}
 
\vfill 
 
\section{Introduction}
Analyzing structure factors and correlations in many-particle systems is a 
standard way to study collective behaviour in real experiments, computer 
simulations and theoretical frameworks. For example, for a system with short 
range microscopic interactions, placed in thermal equilibrium, there are no 
long range spatial correlations in general. Their presence is a typical 
signal that the system is at a critical point. On the other hand, when such 
systems are driven into non-equilibrium steady states, long range 
correlations are often observed \cite{liquids}. Another good example is the  
{\em driven }Ising lattice gas \cite{KLS,BS_RKPZ}. The role of the drive 
(``external field'') is to bias hopping rates along a particular direction 
on the lattice. In addition to many other unexpected features such as 
non-Hamiltonian fixed points controlling the critical behavior \cite{JS_LC}, 
the system exhibits long range spatial correlations {\em at all temperatures  
}above criticality, as a result of the breakdown of the traditional 
fluctuation-dissipation relations \cite{Kubo}. In momentum space, this appears
as a discontinuity singularity of the structure factor at the origin 
\cite{ZHSL}.

In this Paper, we revisit a simple, non-equilibrium model \cite{KSZ2} 
displaying similar features and examine the steady state structure factors 
from a more general point of view. In particular, we will construct {\em the 
full distribution} of various structure factors, using both Monte Carlo 
simulations and Langevin equations following the method in ref\cite{RZ}. The 
model can be considered as a generalization of the KLS \cite{KLS} model, in 
a fashion similar to an equilibrium Ising model being extended to spin-1  
\cite{BEG} or Potts \cite{Potts} models, in that it involves {\em two }%
species of particles (and holes). Referred to as $+$'s and $-$'s, they are 
driven in opposite directions, subject to periodic boundary conditions.
To keep the model simple, we suppress the 
usual Ising nearest neighbor interaction and retain only the excluded 
volume constraint and the bias. In the simplest scenario, charge exchange ($%
+\leftrightarrow -$) is not allowed. Phase transitions in this multi-species 
model, in both one and two dimensions, have been investigated extensively%
\cite{SHZ,VZS,FG,KSZ,traffic,obc}. Monte Carlo simulations \cite{SHZ} in two 
dimensions and mean-field studies \cite{VZS,FG} show that there is a 
transition, controlled by particle density and drive, from a spatially 
homogeneous (disordered) phase to a charge segregated one, where the 
excluded volume constraint leads to the mutual blocking of particles. We 
will focus on the disordered phase of the system, where we have a sound 
analytic understanding of the dynamics in terms of Langevin equations. 
Performing Monte Carlo simulations and measuring the full distributions of 
several structure factors, we find that the Gaussian approximation is 
extremely effective in predicting these properties of our non-equilibrium 
system. In contrast to the earlier study \cite{RZ}, the structure factors 
here form a $2\times2$ {\em matrix}, due to the presence of {\em two} species of 
particles, and the real and imaginary parts of the off-diagonal elements 
exhibit a characteristic asymmetry. The measurement of these distributions  
can be used to extract considerable information  
about the structure of the underlying linearized equations of motion, 
independent of model details.

\section{The Microscopic Model and the Simulations}
We consider a two dimensional fully periodic lattice with $L\times L$ sites, 
each of which can be empty or occupied by a single particle. Since there are 
two species we need two occupation variables $n_{{\bf x}}^{+}$ and $n_{{\bf x%
}}^{-}$, with $n$ being 0 or 1, depending on whether a positive or negative 
particle is present at site ${\bf x}$. Although we refer to these particles 
as ``charged'', they do not interact via the Coulomb potential. Instead, 
they respond to an external, uniform ``electric'' field $E$, directed along 
a specific lattice axis which we label the $+x_{\parallel }$-direction.  
We restrict ourselves to zero total charge, i.e.,  
$\sum_{{\bf x}}[n_{{\bf x}}^{+}-n_{{\bf x}}^{-}]=0$. In the absence of the 
drive, the dynamics does not distinguish between the different species: both 
types hop randomly to nearest neighbor empty sites, with a rate $\Gamma $. 
The electric field introduces a bias into these rates in such a way that 
jumps {\em against} the force will be exponentially suppressed. During one 
Monte Carlo step $2L^2$ nearest neighbor bonds are chosen randomly. If a 
particle-hole pair is encountered, an exchange takes place with probability $%
W=\Gamma \min \{1,\exp (qE\,\delta x_{\parallel })\}$, where $q=\pm 1$ is 
the charge of the particle and $\delta x_{\parallel }=\pm 1,0$ is the change 
of the $x_{\parallel }$ coordinate of the particle due to the jump. 
 
\begin{figure}[tbp] 
\hspace*{2cm}
\epsfxsize=12cm
\epsfysize=12cm
\epsfbox{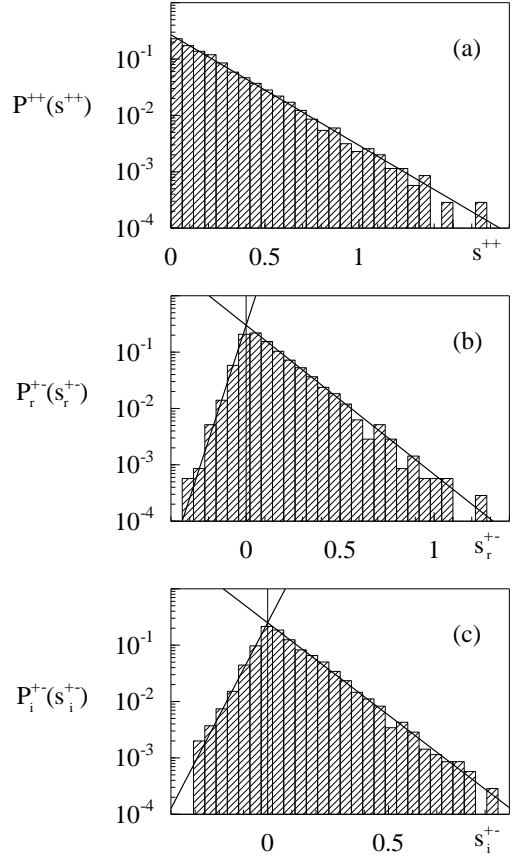}
\caption{Histograms representing the distributions of the ${\bf k}=\frac{2\pi%
}{L}(0,1)$ structure factors for (a) $\frac{ n_{{\bf k}}^{+} n_{-{\bf k}%
}^{+} }{V} $, (b) $\frac{\mbox{Re}[n_{{\bf k}}^{+} n_{-{\bf k}}^{-}] }{V} $ 
and (c) $\frac{\mbox{Im}[n_{{\bf k}}^{+} n_{-{\bf k}}^{-}] }{V} $. $L=60$, $%
E=0.471$ and $\bar{\varrho}=0.175$. Theoretical distributions (a) $%
P^{++}(s^{++};{\bf k})$, (b) $P^{+-}_{r}(s^{+-}_{r};{\bf k})$ and (c) $%
P^{+-}_{i}(s^{+-}_{i};{\bf k})$ are plotted with solid lines on the same 
graphs. } 
\label{fig1} 
\end{figure}

\begin{figure}[tbp]
\hspace*{2cm}
\epsfxsize=12cm
\epsfysize=12cm
\epsfbox{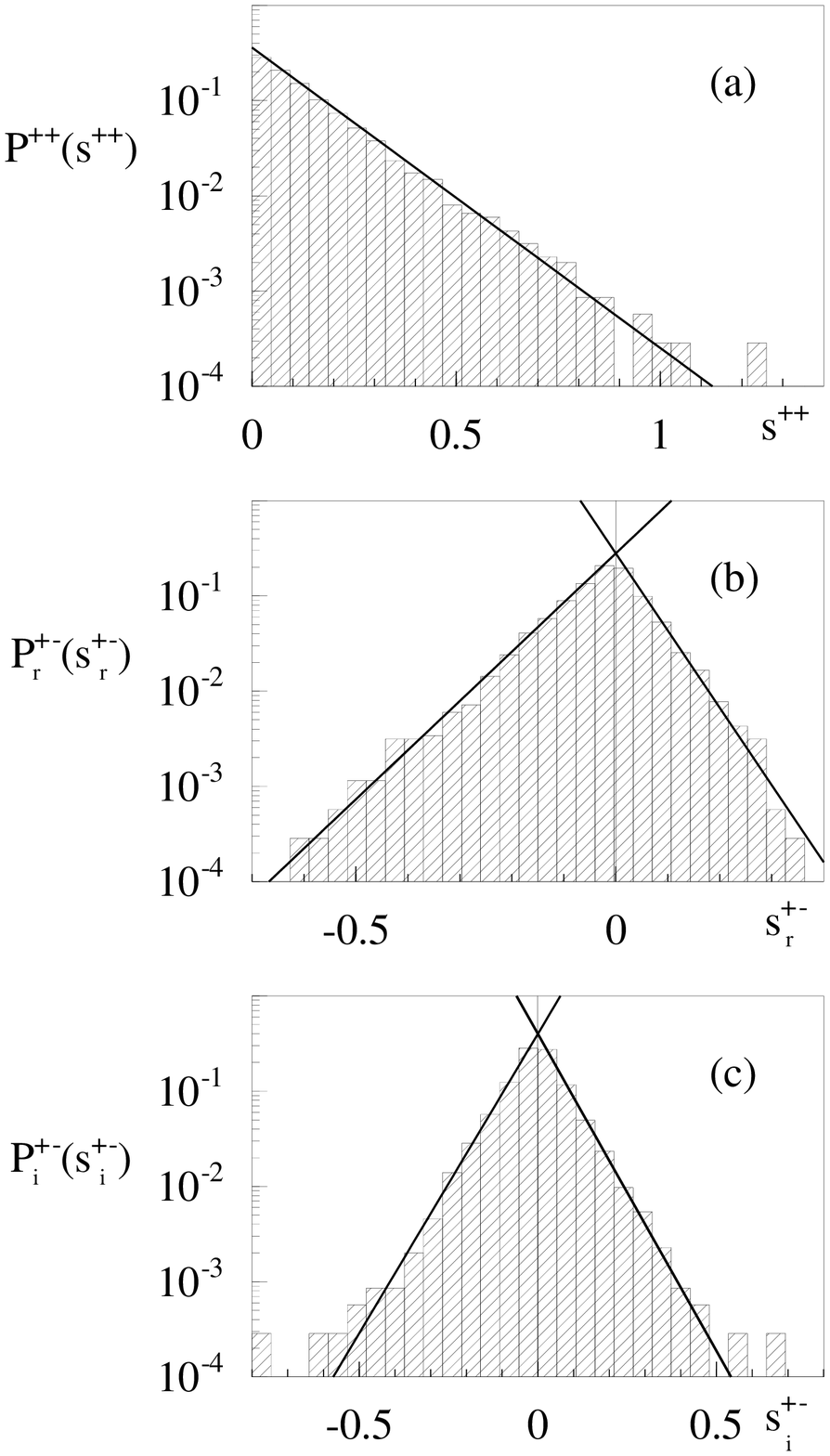} 
\caption{Histograms representing the distributions of the ${\bf k}=\frac{2\pi%
}{L}(1,0)$ structure factors for (a) $\frac{ n_{{\bf k}}^{+} n_{-{\bf k}%
}^{+} }{V} $, (b) $\frac{\mbox{Re}[n_{{\bf k}}^{+} n_{-{\bf k}}^{-}] }{V} $ 
and (c) $\frac{\mbox{Im}[n_{{\bf k}}^{+} n_{-{\bf k}}^{-}] }{V} $. $L=60$, $%
E=0.471$ and $\bar{\varrho}=0.175$. Theoretical distributions (a) $%
P^{++}(s^{++};{\bf k})$, (b) $P^{+-}_{r}(s^{+-}_{r};{\bf k})$ and (c) $%
P^{+-}_{i}(s^{+-}_{i};{\bf k})$ are plotted with solid lines on the same 
graphs. } 
\label{fig2} 
\end{figure}

For our simulations, we set $\Gamma =1$. Using $60\times 60$ lattices, we 
initialize the system with random configurations of various particle 
densities and carry out runs ranging from $2.5$ to $5\times 10^5$ MCS. 
After allowing $62500$ MCS for the system to settle into a steady state, we 
measure the Fourier transforms of $n_{{\bf x}}^{\pm }$ every $125$ MCS, 
defining them in the usual way:  
\begin{equation} 
n_{{\bf k}}^{\pm }=\sum_{{\bf x}}e^{-i\,{\bf kx}}\;n_{{\bf x}}^{\pm }\; 
\label{ftr_sim} 
\end{equation} 
In the literature, the term ``structure factor'' typically refers to the 
(ensemble- or time-) average of density-density operators in momentum space, 
i.e.,  
\begin{equation} 
S^{\alpha \beta }({\bf k})\equiv \frac{1}{V}
\langle n_{{\bf k}}^\alpha n_{-{\bf k}}^\beta \rangle \;, 
\end{equation} 
where $\alpha ,\beta =+,-$ ; ${\bf k}=\frac{2\pi }L(m_{\perp },m_{\parallel 
})$ and $V=L^2$ is the volume. Instead of the overall averages, we construct 
histograms for $\frac{n_{{\bf k}}^{+}n_{-{\bf k}}^{+}}V$, $\frac{\mbox{Re}%
[n_{{\bf k}}^{+}n_{-{\bf k}}^{-}]}V$ and $\frac{\mbox{Im}[n_{{\bf k}}^{+}n_{-%
{\bf k}}^{-}]}V$ from their time series in the steady state. We will use the 
somewhat loose term ``structure factors'' for these fluctuating quantities 
themselves. By symmetry, $\frac{n_{{\bf k}}^{+}n_{-{\bf k}}^{+}}V$ and $%
\frac{n_{{\bf k}}^{-}n_{-{\bf k}}^{-}}V$ are distributed identically, and we 
consider only the former. The results are presented in Figure \ref{fig1} and  
\ref{fig2} for the smallest longitudinal and transverse wave vectors. Before 
discussing the data, we will first present the theoretical 
framework within which they can be understood. 
 
\section{Coarse-grained Description}
Being interested in the large distance (small ${\bf k}$) behavior, we may 
exploit the simplest approach, i.e., to use coarse-grained equations of 
motions for the slowly varying local densities $\varrho ^{\pm }({\bf x},t).$ 
These can be considered as the naive continuum limits of the average lattice 
occupation numbers $\langle n_{{\bf x}}^{\pm }\rangle $ \cite{KSZ2,KSZ}. For 
generality, we consider the d-dimensional case when ${\bf x}_{\parallel }$ 
is directed along the electric field and ${\bf x}_{\perp }$ lies in the $d-1$ 
dimensional subspace transverse to the field. Then the mean-field equations 
of motion for the charge densities follow as:  
\begin{equation} 
\partial _t\varrho ^{\pm }=-\mbox{\boldmath $\nabla$}{\bf \Gamma }\{\varrho 
^{\pm }\stackrel{\leftrightarrow }{\mbox{\boldmath $\nabla$}}(1-\varrho 
^{+}-\varrho ^{-})\pm \varepsilon \hat{{\bf x}}_{\parallel }\varrho ^{\pm 
}(1-\varrho ^{+}-\varrho ^{-})\};\,  \label{meaf} 
\end{equation} 
Here, ${\bf \Gamma }$ is the diffusion matrix which is diagonal but
not a simple multiple 
of the unit matrix, due to the anisotropies induced by the bias. $\stackrel{%
\leftrightarrow }{\mbox{\boldmath $\nabla$}}$ is the asymmetric gradient 
operator, acting on any two functions $f$ and $g$ according to $f\stackrel{%
\leftrightarrow }{\mbox{\boldmath $\nabla$}}g=f\mbox{\boldmath 
$\nabla$}g-g\mbox{\boldmath $\nabla$}f$. $\varepsilon $ is the 
coarse-grained bias and $\hat{{\bf x}}_{\parallel }$ is the unit vector 
along the $x_{\parallel }$ direction. The coefficients appearing in  
Eqn. (\ref{meaf}) are functions of the microscopic control parameters; 
however, their precise dependence need not be known.

To study small fluctuations in the disordered phase, we linearize (\ref{meaf}%
) about the homogeneous time-independent solution $\bar{\varrho}$ and add 
conserved noise terms, to model the fast degrees of freedom. Writing $%
\varrho ^{\pm }({\bf x},t)=\bar{\varrho}+\chi ^{\pm }({\bf x},t)$, we arrive 
at the Langevin equations for the density fluctuations:  
\begin{equation} 
\partial _t\chi ^\alpha ({\bf x},t)={\cal L}^{\alpha \beta }(%
\mbox{\boldmath 
$\nabla$})\chi ^\beta ({\bf x},t)-\mbox{\boldmath $\nabla \eta $}^\alpha (%
{\bf x},t)\;,  \label{rsl} 
\end{equation} 
where  
\begin{equation} 
({\cal L}^{\alpha \beta }(\mbox{\boldmath $\nabla$}))=\left( \matrix{ 
(1-\bar{\varrho})\mbox{\boldmath $\nabla\Gamma\nabla$} - (1-3\bar{\varrho}) 
\varepsilon\Gamma_{\parallel}\partial_{\parallel} & 
\bar{\varrho}\mbox{\boldmath $\nabla\Gamma\nabla$} + \bar{\varrho} 
\varepsilon\Gamma_{\parallel}\partial_{\parallel} \cr 
\bar{\varrho}\mbox{\boldmath $\nabla\Gamma\nabla$} - \bar{\varrho} 
\varepsilon\Gamma_{\parallel}\partial_{\parallel} & 
(1-\bar{\varrho})\mbox{\boldmath $\nabla\Gamma\nabla$} + (1-3\bar{\varrho}) 
\varepsilon\Gamma_{\parallel}\partial_{\parallel} }\right) 
\label{det_matr} 
\end{equation} 
and summation over repeated indices is implied in (\ref{rsl}) 
and in the  
following.
Here, $\mbox{\boldmath $\eta $}^{\pm }({\bf x},t)$ are Gaussian white noise 
terms, satisfying:
\begin{eqnarray}
\begin{array}{rcl}
\langle \eta _i^\alpha ({\bf x},t)\rangle & = & 0 \;,  \\
\;\;\langle \eta _i^\alpha ({\bf x},t)\eta _j^\beta ({\bf x}^{\prime
},t^{\prime })\rangle & = & 2\delta ^{\alpha \beta }\sigma _{ij}\delta ({\bf x}-
{\bf x}^{\prime })\delta (t-t^{\prime }) \;,
\end{array}
\label{rsnoise}
\end{eqnarray} 
where $\alpha ,\beta =+,-$ ; $i,j=1,2,\ldots d$ and $\delta^{\alpha \beta }$ 
is the Kronecker-symbol. $(\sigma _{ij})=%
\mbox{\boldmath $\sigma$}$ is the noise matrix, and similar to ${\bf \Gamma }$,
it is diagonal but not proportional to the unit matrix, due to the bias. 
To describe systems 
in thermal equilibrium, the fluctuation-dissipation theorem would require $%
\mbox{\boldmath $\sigma$}\propto {\bf \Gamma }$. In particular, in the 
absence of the drive, we would have $\mbox{\boldmath $\sigma$}=\bar{\rho}(1-2%
\bar{\rho}){\bf \Gamma }$ here. However, when driven, the proportionality is 
not expected to hold, since the diffusion and noise matrices are 
renormalized differently by the drive $\varepsilon $, similar to the driven 
single species case\cite{JS_LC}. Finally, we point out that $%
\mbox{\boldmath 
$\eta $}^{+}$ and $\mbox{\boldmath $\eta $}^{-}$ are uncorrelated, due to 
the fact that charge exchange is not allowed.

Introducing the Fourier transforms for the density fluctuations (and 
similarly for the noise)  
\begin{equation} 
\chi ^{\pm }({\bf k},\omega )=\int dtd^dx\;\chi ^{\pm }({\bf x}%
,t)\;e^{-i(\omega t+{\bf kx})}\;,  \label{ftr_def} 
\end{equation} 
the solution to (\ref{rsl}) is trivial:  
\begin{equation} 
\chi ^\alpha ({\bf k},\omega )=(L^{-1})^{\alpha \beta }\;i{\bf k}%
\mbox{\boldmath $\eta $}^\beta ({\bf k},\omega )\;,  \label{inv_kol} 
\end{equation} 
where $L^{\alpha \beta }={\cal L}^{\alpha \beta }(i{\bf k})-i\omega \delta 
^{\alpha \beta }$. According to the definition of the full dynamic 
structure factors  
\begin{equation} 
S^{\alpha \beta }({\bf k},\omega )\left[ (2\pi )^{d+1}\delta ({\bf k}+{\bf k}%
^{\prime })\delta (\omega +\omega ^{\prime })\right] \equiv \langle \chi 
^\alpha ({\bf k},\omega )\chi ^\beta ({\bf k}^{\prime },\omega ^{\prime 
})\rangle \;,  \label{dyn_str_def} 
\end{equation} 
we can integrate them over $\omega $ to obtain the steady state structure 
factors:  
\begin{equation} 
S^{\alpha \beta }({\bf k})\left[ (2\pi )^d\delta ({\bf k}+{\bf k}^{\prime 
})\right] \equiv \langle \chi ^\alpha ({\bf k},t)\chi ^\beta ({\bf k}%
^{\prime },t)\rangle \;.  \label{str_def} 
\end{equation} 
Note that they are just the Fourier transforms of the usual {\em equal-time} 
correlation functions $\langle \chi ^\alpha ({\bf x},t)\chi ^\beta ({\bf 0}%
,t)\rangle \equiv G^{\alpha \beta }({\bf x})\;.$ The explicit forms of the 
structure factors are:  
\begin{eqnarray} 
S^{++}({\bf k}) &=&\,\frac{(1-\bar{\varrho})}{(1-2\bar{\varrho})}\,
\frac{ {\bf k} \mbox{\boldmath $ \sigma $} {\bf k} }{ {\bf k\Gamma k} }\;
\frac{({\bf k\Gamma k})^2+ 
\frac{(1-3\bar{\varrho})^2}{(1-\bar{\varrho})^2}\Gamma _{\parallel 
}^2\varepsilon ^2k_{\parallel }^2}{({\bf k\Gamma k})^2+(1-4\bar{\varrho}
)\Gamma _{\parallel }^2\varepsilon ^2k_{\parallel }^2}  \nonumber \\ 
\mbox{Re}\{S^{+-}({\bf k})\} &=&-\,\frac{\bar{\varrho}}{(1-2\bar{\varrho})}\, 
\frac{ {\bf k} \mbox{\boldmath $ \sigma $} {\bf k} }{ {\bf k\Gamma k} }\;
\frac{({\bf k\Gamma k} 
)^2-\frac{(1-3\bar{\varrho})}{(1-\bar{\varrho})}\Gamma _{\parallel 
}^2\varepsilon ^2k_{\parallel }^2}{({\bf k\Gamma k})^2+(1-4\bar{\varrho}
)\Gamma _{\parallel }^2\varepsilon ^2k_{\parallel }^2}  \label{exp_str} \\ 
\mbox{Im}\{S^{+-}({\bf k})\} &=&\,\frac{2\bar{\varrho}}{(1-\bar{\varrho})}\;
\frac{ {\bf k}\mbox{\boldmath $ \sigma $}{\bf k}\,\Gamma_{\parallel}\varepsilon 
k_{\parallel }}{({\bf k\Gamma k})^2+(1-4\bar{\varrho})\Gamma _{\parallel 
}^2\varepsilon ^2k_{\parallel }^2}\;,  \nonumber 
\end{eqnarray}
where \mbox{${\bf k}=({\bf k}_{\perp},k_{\parallel})$}. 
They clearly exhibit a finite discontinuity singularity at the origin, i.e.,  
$\lim_{k_{\parallel}\rightarrow 0}S^{\alpha \beta }({\bf 0},k_{\parallel})\neq 
\lim_{{\bf k}_{\perp}\rightarrow {\bf 0}}S^{\alpha \beta}({\bf k}_{\perp},0)$ 
which translates to power law decays in the spatial 
correlations. In particular, we find the expected \cite{powers} power law, $%
r^{-d}$, related to $\mbox{\boldmath $\sigma$}\not\propto{\bf \Gamma }$, 
typical of non-equilibrium steady states of a system with anisotropy and 
subjected to a conservation law. In addition, a novel power,  
$r^{-(d+1)/2}$  
\cite{KSZ2}, controls correlations along the bias direction. 
 
Next, we seek the {\em full probability distributions} of the 
density-density operators. Thus, for each ${\bf k}$ vector, we construct the 
distributions separately for $\frac{\chi ^{+}({\bf k},t)\chi ^{+}(-{\bf k},t)%
}V$, $\frac{\mbox{Re}[\chi ^{+}({\bf k},t)\chi ^{-}(-{\bf k},t)]}V$ and $%
\frac{\mbox{Im}[\chi ^{+}({\bf k},t)\chi ^{-}(-{\bf k},t)]}V$, represented 
by $s^{++}$, $s_r^{+-}$ and $s_i^{+-}$:  
\begin{eqnarray} 
P^{++}(s^{++};{\bf k}) &=&\left\langle \delta \left( \frac{\chi ^{+}({\bf k}%
,t){\chi ^{+}}^{*}({\bf k},t)}V-s^{++}\right) \right\rangle   \nonumber \\ 
P_r^{+-}(s_r^{+-};{\bf k}) &=&\left\langle \delta \left( \frac{\mbox{Re}%
[\chi ^{+}({\bf k},t){\chi ^{-}}^{*}({\bf k},t)]}V-s_r^{+-}\right) 
\right\rangle   \label{prob_distr} \\ 
P_i^{+-}(s_i^{+-};{\bf k}) &=&\left\langle \delta \left( \frac{\mbox{Im}%
[\chi ^{+}({\bf k},t){\chi ^{-}}^{*}({\bf k},t)]}V-s_i^{+-}\right) 
\right\rangle \;.  \nonumber 
\end{eqnarray} 
The normalization by $V$ helps to avoid infinities in the expectation values 
of $s^{++},s_r^{+-},s_i^{+-}$ as can be seen from (\ref{str_def}) and by 
noting that $(2\pi )^d\delta ({\bf k}\!=\!{\bf 0})=V$ in the infinite volume 
limit. Also note that we used $\chi ^{\pm }(-{\bf k},t)={\chi ^{\pm }}^{*}(%
{\bf k},t)$, since the densities ${\chi ^{\pm }}({\bf r},t)$ are real. For 
our purposes, we need both, the density fluctuations and the explicit 
distribution of the noise, in the $({\bf k},t)$ domain. Within the linear 
stability regime of the disordered phase, each matrix element of $\left( 
(L^{-1})^{\alpha \beta }\right) $ has two poles in the positive $\omega $ 
half-plane corresponding to two stable eigenvalues of $({\cal L}^{\alpha 
\beta })$. Thus, from (\ref{inv_kol}) we find:  
\begin{equation} 
\chi ^\alpha ({\bf k},t)=\int_{-\infty }^tdt^{\prime }\;\left[ \int_{-\infty 
}^\infty \frac{d\omega }{2\pi }e^{i\omega (t-t^{\prime })}(L^{-1})^{\alpha 
\beta }\right] \;i{\bf k}\mbox{\boldmath $\eta $}^\beta ({\bf k},t^{\prime 
})\;  \label{inv_ktl} 
\end{equation} 
with  
\begin{equation} 
P[\eta _i^\alpha ({\bf k},t)]\propto exp\left\{ -\frac 12\int dtd^dk\;\eta 
_i^\alpha ({\bf k},t)\frac{(D^{-1})_{ij}^{\alpha \beta }}{(2\pi )^d}{\eta 
_j^\beta }^{*}({\bf k},t)\right\} \;,  \label{kt_noise} 
\end{equation} 
where $D_{ij}^{\alpha \beta }\equiv 2\delta ^{\alpha \beta }\sigma _{ij}$ in 
our model. Due to translational invariance, fields with different ${\bf k}$ 
vectors are decoupled, so we will suppress ${\bf k}$ in the following. Then (%
\ref{inv_ktl}) can be written as  
\begin{equation} 
\chi ^\alpha (t)=v_j^{\alpha \beta }(t,t^{\prime })\;\eta _j^\beta 
(t^{\prime })  \label{fluct_noise} 
\end{equation} 
where  
\begin{equation} 
v_j^{\alpha \beta }(t,t^{\prime })=\Theta (t-t^{\prime })\;\left[ 
\int_{-\infty }^\infty \frac{d\omega }{2\pi }e^{i\omega (t-t^{\prime 
})}(L^{-1})^{\alpha \beta }\right] \;ik_j\;.  \label{vmatr} 
\end{equation} 
Note that summation over repeated indices also includes an integral over 
$t^{\prime }$ in (\ref{fluct_noise}). 
 
We start with the probability distribution of $\frac{\chi ^{+}(t){\chi ^{+}}%
^{*}(t)}V$ by first finding its characteristic function, i.e.,  
\begin{equation} 
\tilde{P}^{++}(\Omega )=\int_{-\infty }^\infty ds^{++}\;e^{i\Omega 
s^{++}}P^{++}(s^{++})=\left\langle e^{i\Omega \frac{\chi ^{+}(t){\chi ^{+}}%
^{*}(t)}V}\right\rangle \;.  \label{char++} 
\end{equation} 
When performing the average in (\ref{char++}), all integrations over the 
noise are trivial, except those associated with $\pm {\bf k}$. Thus, we need 
to evaluate the following integral:  
\begin{equation} 
\tilde{P}^{++}(\Omega )=\int \prod_{t^{\prime},\gamma ,j}d\eta _j^\gamma 
(t^{\prime})\,d{\eta _j^\gamma }^{*}(t^{\prime})\;
P[\eta _i^\alpha (t),{\eta _i^\alpha }%
^{*}(t)]\;e^{i\Omega \frac{\chi ^{+}(t){\chi ^{+}}^{*}(t)}V}  \label{char++1} 
\end{equation} 
where $\eta _j^\gamma (t^{\prime})$ and ${\eta _j^\gamma }^{*}(t^{\prime})$ 
are the ${\bf k}$ 
and $-{\bf k}$ components of the noise, respectively, yielding the only 
non-trivial integrations. Inserting (\ref{fluct_noise}) into (\ref{char++1}%
), we still have a Gaussian integrand, controlled by the quadratic form:  
\begin{equation} 
\eta _\mu \left( (D^{-1})_{\mu \nu }-i\Omega v_\mu ^{+}{v_\nu ^{+}}%
^{*}\right) {\eta _\nu }^{*}\;,  \label{quadr++} 
\end{equation} 
where the indices $\mu ,\nu $ include all the degrees of freedom left over, 
namely, time, charge, and spatial component. The corresponding path 
integrals lead to  
\begin{equation} 
\tilde{P}^{++}(\Omega )=\frac{\mbox{det}\left[ (D^{-1})_{\mu \nu 
}\right] }{\mbox{det}\left[ (D^{-1})_{\mu \nu }-i\Omega v_\mu ^{+}{v_\nu ^{+}%
}^{*}\right] }=\frac 1{\mbox{det}\left[ \delta _{\mu \nu }-i\Omega D_{\mu 
\gamma }v_\gamma ^{+}{v_\nu ^{+}}^{*}\right] }\;,  \label{det++} 
\end{equation} 
where the numerator in the middle expression originates in the normalization 
factor ensuring $\tilde{P}^{++}(0)=1$. Exploiting the formula $\mbox{det}%
\left[ \delta _{\mu \nu }+a_\mu b_\nu \right] =1+a_\mu b_\mu ,$ we obtain  
\begin{equation} 
\tilde{P}^{++}(\Omega )=\frac 1{1-i\Omega v_\mu ^{+}D_{\mu \nu }{v_\nu 
^{+}}^{*}}\;.  \label{char++2} 
\end{equation} 
Note that the coefficient of $i\Omega $ is simply the `$++$' structure 
factor:  
\begin{equation} 
v_\mu ^{+}D_{\mu \nu }{v_\nu ^{+}}^{*}=\left\langle \frac{\chi ^{+}({\bf k}%
,t){\chi ^{+}}^{*}({\bf k},t)}V\right\rangle =S^{++}({\bf k})\;. 
\label{trick++} 
\end{equation} 
Taking the inverse transform to obtain $P^{++}$, the single pole $-i/S^{++}(%
{\bf k})$ in the lower half $\Omega -$plane yields an exponential 
distribution for the non-negative variable $s^{++}$, i.e.,  
\begin{eqnarray} 
P^{++}(s^{++};{\bf k})=\left\{  
\begin{array}{cc} 
\frac 1{S^{++}({\bf k})}\;e^{-s^{++}/S^{++}({\bf k})} & \;\;\mbox{if}%
\;\;s^{++}\geq 0 \\  
0 & \;\;\mbox{if}\;\;s^{++}<0 
\end{array} 
\right.  \label{prob++} 
\end{eqnarray} 
We will refer to $1/S^{++}({\bf k})$ as the ``decay factor'' of the 
exponential. 
 
Next, we consider the distribution of $\frac{\mbox{Re}[\chi ^{+}(t){\chi ^{-}%
}^{*}(t)]}V$. Following the same steps, we see that the quadratic form 
controlling the Gaussian integrand is  
\begin{equation} 
\eta _\mu \left( (D^{-1})_{\mu \nu }-\frac{i\Omega }2(v_\mu ^{+}{v_\nu ^{-}}%
^{*}+v_\mu ^{-}{v_\nu ^{+}}^{*})\right) {\eta _\nu }^{*}\;,  \label{quadr+-} 
\end{equation} 
leading to  
\begin{equation} 
\tilde{P_r}^{+-}(\Omega )=\frac 1{\mbox{det}\left[ \delta _{\mu \nu }-%
\frac{i\Omega }2(D_{\mu \gamma }v_\gamma ^{+}{v_\nu ^{-}}^{*}+D_{\mu \gamma 
}v_\gamma ^{-}{v_\nu ^{+}}^{*})\right] }\;.  \label{det+-} 
\end{equation} 
Using the formula $\mbox{det}\left[ \delta _{\mu \nu }+a_\mu b_\nu +c_\mu 
d_\nu \right] =1+[a_\mu b_\mu +c_\mu d_\mu ]+[(a_\mu b_\mu )(c_\nu d_\nu 
)-(a_\mu d_\mu )(b_\nu c_\nu )]$ we arrive at  
\begin{equation} 
\tilde{P_r}^{+-}(\Omega )=\frac 1{1-i\Omega \;\mbox{Re}[S^{+-}({\bf k})]+%
\frac{\Omega ^2}4[|S^{++}({\bf k})|^2-|S^{+-}({\bf k})|^2]}\;, 
\label{char+-} 
\end{equation} 
where, similarly to (\ref{trick++}), we have used  
\begin{equation} 
v_\mu ^{+}D_{\mu \nu }{v_\nu ^{-}}^{*}=\left\langle \frac{\chi ^{+}({\bf k}%
,t){\chi ^{-}}^{*}({\bf k},t)}V\right\rangle =S^{+-}({\bf k})\;. 
\label{trick+-} 
\end{equation} 
Unlike the previous case, $\tilde{P_r}^{+-}(\Omega )$ has two poles: one 
($\Omega _{-}$) being on the negative, and the other ($\Omega _{+}$) on the 
positive, imaginary axis:  
\begin{equation} 
\Omega _{\mp }=\frac{2i}{\Delta} \;\left( \mbox{Re}[S^{+-}({\bf k})] \mp  
\sqrt{\Delta + (\mbox{Re}[S^{+-}({\bf k})])^2} \right)  \label{params} 
\end{equation} 
where $\Delta \equiv |S^{++}({\bf k})|^2-|S^{+-}({\bf k})|^2>0$\ . In 
general, their magnitudes are different. Thus, the inverse transform yields 
an {\em asymmetric} exponential distribution, characterized by two distinct 
decay factors $|\Omega _{+}|$ and $|\Omega _{-}|$:  
\begin{eqnarray} 
P_r^{+-}(s_r^{+-};{\bf k})=\left\{  
\begin{array}{cc} 
\frac {1}{N}\;e^{-|\Omega _{-}|s_r^{+-}} & \;\;\mbox{if}\;\;s_r^{+-}\geq 0 
\\  
\frac {1}{N}\;e^{ |\Omega _{+}|s_r^{+-}} & \;\;\mbox{if}\;\;s_r^{+-}<0 
\end{array} 
\right.  \label{prob+-} 
\end{eqnarray} 
with $N=\sqrt{\Delta + (\mbox{Re}[S^{+-}({\bf k})])^2}$. To obtain the 
distribution of $\frac{\mbox{Im}[\chi ^{+}(t){\chi ^{-}}^{*}(t)]}V$ we 
simply interchange $\mbox{Re}[S^{+-}({\bf k})]$ and $\mbox{Im}[S^{+-}({\bf k}%
)]$ in eqs. (\ref{params},\ref{prob+-}). 
 
One important consequence of (\ref{prob++}) and (\ref{prob+-}) is that their 
standard deviations always take a value {\em greater or equal} than the 
average. In particular  
\begin{eqnarray} 
\sqrt{\overline{(s^{++})^2}-\left( \overline{s^{++}}\right) ^2} &=&S^{++}(%
{\bf k})  \nonumber \\ 
\sqrt{\overline{(s_r^{+-})^2}-\left( \overline{s_r^{+-}}\right) ^2} &=&\sqrt{%
(\mbox{Re}[S^{+-}({\bf k})])^2+\frac \Delta 2}  \label{stdev} \\ 
\sqrt{\overline{(s_i^{+-})^2}-\left( \overline{s_i^{+-}}\right) ^2} &=&\sqrt{%
(\mbox{Im}[S^{+-}({\bf k})])^2+\frac \Delta 2}\;.  \nonumber 
\end{eqnarray} 
Thus, when measuring structure factors in the {\em disordered} phase, 
fluctuations comparable to the average should not come as a surprise. We 
also emphasize that formulas (\ref{prob++}) and (\ref{prob+-}) are 
completely independent of the specific model and can be derived in any 
multicomponent system. The only necessary conditions are linear Langevin 
equations with Gaussian noise. 

\section{Discussion and Summary} 
Finally, let us turn to a comparison with the simulation results summarized 
in Figure \ref{fig1} and \ref{fig2}, for the two smallest wave vectors. The 
`$++$' histograms show the usual exponential decay \cite{RZ}, while the `$+-$' 
histograms clearly represent asymmetric exponential distributions. To test 
the predictions of our Gaussian theory, namely, that the slopes of the 
histograms are determined by the structure factor averages themselves, we 
simply measured these averages, i.e., $S^{++}$ , $\mbox{Re}S^{+-}$ and $%
\mbox{Im}S^{+-}$. The `$++$' case is particularly simple since the decay 
factor is just the inverse of $S^{++}$ itself. For the two '$+-$' 
distributions, we inserted the {\em measured} averages into the non-trivial  
{\em theoretical} relationship (\ref{params}), to find the decay factors $%
|\Omega _{\mp }|$. The agreement between simulation results and theory is 
remarkably good, indicating that linearized Langevin equations are quite 
acceptable in this regime. While the external field, $E$, may obviously generate 
renormalizations, these can be absorbed in the effective parameters of the 
theory, namely the diffusion matrix, the noise matrix, the average density 
and the coarse-grained bias, leaving the {\em form} of the structure factor 
distributions invariant. However, we must avoid critical fluctuations since 
the linear approximation will fail to capture their effects correctly. 
 
To summarize, we have analyzed the full structure factor distributions in a 
simple driven lattice gas. Due to the presence of two species, there are 
several distinct structure factor distributions here. All are simple 
exponentials, with the decay factors given in terms of the average structure 
factors. However, only the ``diagonal'' elements $P^{++}$ and $P^{--}$ are 
characterized by a {\em single} decay factor, being distributions for a 
non-negative variable. In contrast, both $P_r^{+-}$ and $P_i^{+-}$ are 
combinations of {\em two} exponentials, with {\em different} decay factors, 
characterizing positive versus negative arguments, respectively. While this 
asymmetry may, at first sight, be rather surprising, its origin is quite 
easily traced to the presence of {\em off-diagonal} elements in the matrix $%
{\cal L}^{\alpha \beta }$, Eqn. (\ref{det_matr}). Thus, the observation of 
asymmetric structure factor histograms, either in an experiment or a 
simulation, provides some basic insight into the form of the underlying 
Langevin equation. 
 
\section*{Acknowledgements} 
 
We thank Z. Toroczkai, S. Sandow and C. Laberge for many stimulating 
discussions. This research is supported in part by grants from the National 
Science Foundation through the Division of Materials Research .

\newpage 
 
\end{document}